\documentclass[aps,prx,reprint]{revtex4-2}
\usepackage{mathtools}

\onecolumngrid
\usepackage{amsmath}
\usepackage{graphicx}
\usepackage{placeins}
\usepackage{float}
\usepackage{subfigure}
\usepackage{dcolumn}
\usepackage{bm}

\begin{document}


\title{Experimental Blueprint for Distinguishing Decoherence from Objective Collapse}
\author{Ridha Horchani}
\email{horchani@squ.edu.om}
 \affiliation{Department of Physics, College of Science, Sultan Qaboos University, P.O. Box 36, P.C. 123, Al-Khod, Muscat, Sultanate of Oman}

\begin{abstract}
The transition from the quantum to the classical realm remains one of the most profound open questions in physics.
While quantum theory predicts the existence of macroscopic superpositions, their apparent absence in the everyday world
is attributed either to environmental decoherence or to an intrinsic mechanism for wave-function collapse.
This work presents a quantitative and experimentally grounded framework for distinguishing these possibilities.
We propose a levitated-optomechanical platform capable of generating controllable Schrödinger-cat states in the center-of-mass motion of a dielectric nanosphere.  A comprehensive master equation incorporates gas collisions, black-body radiation, and photon-recoil noise, establishing a calibrated environmental baseline.
The Continuous Spontaneous Localization (CSL) model is embedded within the same framework,
predicting a characteristic saturation of the decoherence rate with superposition size and a quadratic scaling with mass.
A Bayesian inference protocol is outlined to discriminate collapse-induced excess decoherence from environmental noise.
Together these elements provide a concrete experimental blueprint for a decisive test of quantum linearity,
either revealing new physics beyond standard quantum mechanics or setting the most stringent bounds to date on objective-collapse parameters.\\
\textbf{keywords:} Quantum Foundations, Macroscopic Superposition, Wavefunction Collapse, Continuous Spontaneous Localization (CSL), Levitated Optomechanics, Decoherence, Master Equation.
\end{abstract}
\maketitle
\section{Introduction}
The mathematical framework of quantum mechanics is among the most precisely verified theories in all of science. From atomic spectra to the behavior of subatomic particles, its predictions have been confirmed with extraordinary accuracy. Yet, a profound conceptual tension persists between its axioms and classical experience. Quantum superposition and entanglement allow a single object to exist simultaneously in multiple states or locations, starkly contrasting the definite outcomes of the macroscopic world~\cite{dirac1930quantum, ballentine2014quantum}.
This quantum–classical paradox was epitomized by Schrödinger’s famous thought experiment, in which a cat exists simultaneously alive and dead~\cite{schrodinger1935cat}. The Copenhagen interpretation resolves this by invoking a non-unitary wavefunction collapse during measurement, but this solution only shifts the problem to defining what constitutes a “measurement.” Modern quantum theory addresses this through \textit{decoherence}~\cite{zurek2003decoherence, joos2013decoherence}, which demonstrates how environmental interactions suppress interference, transforming quantum superpositions into classical mixtures. However, decoherence alone does not solve the measurement problem: it explains the disappearance of interference but not the emergence of a single, definite outcome, leaving the universal wavefunction in a superposition~\cite{bell1987speakable, zurek2018quantum}.\\
This limitation has motivated alternative approaches, notably \textit{objective collapse models}~\cite{bassi2013models, ghirardi1986unified, pearle1989combining}. These models posit small stochastic and nonlinear modifications to the Schrödinger equation, causing the wavefunction of sufficiently massive objects to spontaneously localize. In the influential Continuous Spontaneous Localization (CSL) model~\cite{ghirardi1990markov, bassi2003dynamical}, a universal noise field induces collapses at a rate proportional to mass, ensuring microscopic quantum behavior while enforcing macroscopic classicality.\\
Recent experimental progress has brought this quantum–classical boundary into direct empirical reach. Matter-wave interferometry has demonstrated superposition with increasingly large molecules, from $\mathrm{C}_{60}$ fullerenes~\cite{arndt1999wave} to organic complexes exceeding $25{,}000$~amu~\cite{eibenberger2013matter, fein2019quantum}. Concurrently, cavity optomechanics has achieved quantum ground-state cooling~\cite{chan2011laser} and entanglement between mechanical oscillators and light~\cite{palomaki2013entangling}. A particularly promising platform is levitated optomechanics, where nanoparticles are isolated in ultra-high vacuum using optical or electromagnetic traps~\cite{romeroisart2011optically, Millen2020, Vinante2023, toros2020}. This extreme isolation enables the preparation of massive objects in quantum states, creating an unprecedented opportunity to test collapse models.\\
Building on these developments, recent experiments have placed increasingly stringent bounds on collapse dynamics, yet key limitations remain. For instance, Vinante \textit{et al.}~\cite{Vinante2023} set bounds on the CSL rate using a cryogenic, magnetically levitated nanoparticle, but their approach relied on indirect noise spectroscopy. \\
A decisive test, therefore, must not only create large superpositions but also quantitatively account for all known environmental decoherence to identify any residual excess as potential evidence of collapse. This work provides a unified theoretical blueprint for such an experiment, based on an optically levitated dielectric nanosphere coupled to an internal two-level system, enabling deterministic creation of macroscopic superpositions via a conditional displacement gate. Our master-equation model incorporates gas collisions, blackbody radiation, and trap noise to establish a precise environmental baseline. By embedding the CSL mechanism within this framework, we derive distinctive testable signatures—namely, saturation with superposition size and quadratic mass scaling—hallmarks of objective collapse.\\
Unlike previous analyses that treated environmental decoherence and stochastic localization separately, this study establishes a unified, fully calibratable framework that integrates both effects within a single master equation and provides a Bayesian inference protocol for model discrimination. This method transforms the test from indirect parameter bounding into a quantitative hypothesis comparison, thereby closing the metrological loop between theory and experiment. This work thus transforms the century-old measurement problem from a philosophical paradox into a concrete, falsifiable experimental question, offering a clear path toward resolving whether classicality emerges from environmental entanglement or a fundamental modification of quantum mechanics.
\section{Theoretical Framework for Macroscopic Superposition Generation and Decoherence Analysis}
This section outlines the complete theoretical model for creating a spatial superposition of a levitated nanosphere and distinguishing environmental decoherence from potential collapse models. We begin by defining the system and its target state, then detail the quantum control protocol to generate it. Subsequently, we model all dominant decoherence channels and finally present the data analysis framework for testing wavefunction collapse theories.
\subsection{System Definition and Cat State Generation via Conditional Displacement}
We consider a dielectric nanosphere of mass \(m\) and radius \(R\), whose center-of-mass (COM) motion along one axis (\(x\)) is confined by a high-frequency optical dipole trap. This system is described by the harmonic oscillator Hamiltonian:
\begin{equation}
\hat H_0 = \frac{\hat p^2}{2m} + \frac{1}{2}m\Omega^2 \hat x^2 = \hbar\Omega\!\left(\hat a^\dagger \hat a + \frac{1}{2}\right),
\end{equation}
where \(\Omega\) is the trap frequency, and \(\hat a^\dagger\) (\(\hat a\)) are the creation (annihilation) operators. The COM position and momentum operators are expressed in terms of the zero-point fluctuation amplitude, \(x_{\text{zpf}} = \sqrt{\frac{\hbar}{2m\Omega}}\), as:
\begin{equation}
\hat x = x_{\text{zpf}}(\hat a + \hat a^\dagger), \qquad \hat p = \frac{i\hbar}{2x_{\text{zpf}}}(\hat a^\dagger - \hat a).
\end{equation}
The target quantum state is a spatial superposition, or "cat state," \(|\psi_{\text{cat}}\rangle = \frac{1}{\sqrt{\mathcal N}} (|+\alpha\rangle + |-\alpha\rangle)\), where \(|\pm\alpha\rangle\) are coherent states and \(\mathcal N = 2(1 + e^{-2|\alpha|^2})\) is a normalization factor. The physical separation between the two superposed wavepackets is \(\Delta x = 2|\alpha|x_{\text{zpf}}\), which represents the macroscopicity of the superposition.\\
To generate this state, we employ an internal two-level system (TLS), such as a defect center, within the nanosphere, described by Pauli operators \(\hat\sigma_x,\hat\sigma_y,\hat\sigma_z\). A short, off-resonant optical pulse induces an AC-Stark shift, described by the interaction Hamiltonian:
\begin{equation}
\hat H_{\text{int}} = \hbar \chi(t)\, \hat\sigma_z \cos(k\hat x+\phi),
\end{equation}
where \(k\) is the optical wavevector and \(\hat\sigma_z\) is the Pauli operator. By setting the phase \(\phi=\pi/2\) and assuming a pulse duration much shorter than the oscillator period (\(\tau\!\ll\!2\pi/\Omega\)), the time evolution results in the exact unitary gate:
\begin{equation}
\hat U_{\text{gate}} = \exp[i\theta\hat\sigma_z \sin(k\hat x)], \quad \theta = 2\!\int_0^\tau\!\chi(t)\,dt .
\end{equation}
In the Lamb-Dicke regime (\(k x_{\text{zpf}}\!\ll\!1\)), we approximate \(\sin(k\hat x)\!\approx\!k\hat x\). This allows us to rewrite the gate as a conditional displacement operation:
\begin{equation}
\hat U_{\text{gate}} = e^{-(\eta\theta)^2/2}\, \hat D\!\big(i\,\eta\theta\,\hat\sigma_z\big), \qquad \eta = k x_{\text{zpf}},
\end{equation}
where \(\hat D(\beta)=\exp(\beta \hat a^\dagger - \beta^* \hat a)\) is the displacement operator. By initializing the TLS in a superposition \((|g\rangle+|e\rangle)/\sqrt2\) and applying the gate, a subsequent projective measurement of the TLS conditionally prepares the motional cat state \(|\psi_{\text{cat}}\rangle \propto |+i\eta\theta\rangle + |-i\eta\theta\rangle\), with a size \(|\alpha| = \eta\theta\) that is tunable via the pulse strength \(\theta\).
\subsection{Comprehensive Decoherence Modeling}
The fragility of macroscopic superpositions necessitates a rigorous model of decoherence. For Markovian processes that cause position localization, the system's density matrix \(\hat\rho\) evolves according to a master equation:
\begin{align}
\partial_t \hat\rho = -\frac{i}{\hbar}[\hat H_0,\hat\rho] -\frac{D_{pp}}{\hbar^2}[\hat x,[\hat x,\hat\rho]] +& \gamma_m(n_{\rm th}\!+\!1)\mathcal D[\hat a]\hat\rho\\ 
& + \gamma_m n_{\rm th}\mathcal D[\hat a^\dagger]\hat\rho \nonumber
\label{eq:master_env}
\end{align}
where \(\mathcal D[\hat L]\hat\rho = \hat L\hat\rho \hat L^\dagger -\frac{1}{2}\{\hat L^\dagger\hat L,\hat\rho\}\) is the Lindblad dissipator modeling thermalization. The key term for spatial decoherence is the one containing the momentum diffusion constant \(D_{pp}\), which aggregates noise from several environmental sources:
\begin{align}
D_{pp}^{\rm gas}  &= \frac{8}{\sqrt{2\pi}}\, \frac{P}{v_T}\frac{m^2 m_g}{(m+m_g)^2}\sigma,\quad \text{(Gas collisions)}\\
& v_T = \sqrt{\frac{2k_B T}{m_g}}  \nonumber \\
D_{pp}^{\rm trap} &= \frac{2}{3}\Gamma_{\rm sc}(\hbar k)^2, \quad \text{(Trap photon shot noise)} \\
D_{pp}^{\rm bb}   &= D_{pp}^{\rm bb,emit} + D_{pp}^{\rm bb,abs} + D_{pp}^{\rm bb,scat}, \quad \text{(Blackbody radiation)}
\end{align}
The decoherence rate of the cat state's off-diagonal matrix element \(\langle \alpha | \hat\rho | -\alpha \rangle\) due to these environmental sources is:
\begin{equation}
\Gamma_{\rm env}(\Delta x) = \frac{D_{pp}}{\hbar^2}\,\Delta x^2 = \frac{4 D_{pp} x_{\text{zpf}}^2}{\hbar^2}\,|\alpha|^2 .
\end{equation}
Crucially, \(D_{pp}\) can be independently calibrated by measuring the oscillator's heating rate \(\dot n\), via \(D_{pp} = \hbar m \Omega \dot n\), \emph{before} any cat state is created.\\
We now integrate the Continuous Spontaneous Localization (CSL) model, a leading wavefunction collapse theory. CSL posits a universal noise field that causes spatial localization at a rate proportional to mass. Its effect is incorporated as an additional term in the master equation:
\begin{equation}
\partial_t\hat\rho = -\frac{i}{\hbar}[\hat H_0,\hat\rho] -\frac{\lambda_{\rm CSL}r_C^3}{(4\pi)^{3/2}m_0^2} \!\int\!d^3q\,e^{-q^2r_C^2/4} [\tilde\mu(\mathbf q),[\tilde\mu(-\mathbf q),\hat\rho]],
\end{equation}
where \(\lambda_{\rm CSL}\) is the collapse rate parameter, \(r_C\approx 100\) nm is the correlation length, \(m_0\) is a reference mass (1 amu), and \(\tilde\mu(\mathbf q)\) is the Fourier-transform of the mass density. For a homogeneous sphere of radius \(R\) and mass \(m\), the resulting CSL-induced decoherence rate is:
\begin{equation}
\Gamma_{\rm CSL}(\Delta x) = \lambda_{\rm CSL}\!\left(\frac{m}{m_0}\right)^2 \!\!\left[ 1 - \frac{\exp\!\big(-\tfrac{\Delta x^2}{4r_C^2(1+R^2/r_C^2)}\big)} {(1+R^2/r_C^2)^{3/2}} \right].
\end{equation}
This formula reveals two key asymptotic regimes: 
\begin{align}
\Gamma_{\rm CSL}(\Delta x\!\ll\!r_C) &\approx \lambda_{\rm CSL}\!\left(\tfrac{m}{m_0}\right)^2 \tfrac{\Delta x^2}{4r_C^2(1+R^2/r_C^2)}, \\
\Gamma_{\rm CSL}(\Delta x\!\gg\!r_C) &\to \Gamma_{\rm CSL}^{\max} = \lambda_{\rm CSL}\!\left(\tfrac{m}{m_0}\right)^2 .
\end{align}
For small superpositions (\(\Delta x\!\ll\!r_C\)), \(\Gamma_{\rm CSL} \propto \Delta x^2\), mimicking environmental decoherence. However, for large superpositions (\(\Delta x\!\gg\!r_C\)), the rate saturates at a maximum value. This saturation and its quadratic mass scaling are unique fingerprints of CSL, distinct from the ever-increasing \(\Gamma_{\rm env} \propto \Delta x^2\). \\
For comparison, the Diósi-Penrose (DP) gravitational collapse model predicts a rate:
\begin{equation}
\Gamma_{\rm DP}(\Delta x) = \frac{G m^2}{\hbar \sqrt{\pi} R_0}\left(1-\frac{R_0}{\sqrt{R_0^2+\Delta x^2}}\right),
\end{equation}
where \(R_0\!\sim\!10^{-10}\,\mathrm{m}\) is the self-gravitational cutoff radius. Unlike CSL, the DP rate lacks saturation and scales inversely with \(\Delta x\) for large separations, providing a further discriminant.
\subsection{Experimental Feasibility and Parameter Regimes}
\label{subsec:feasibility}
While the theoretical framework developed above is general, the
realization of a decisive collapse test demands quantitative
environmental suppression.  In Table \ref{tab:parameters} we estimate realistic parameter windows in which environmental decoherence becomes comparable to the
target CSL signal.\\
For a dielectric nanosphere of radius $R = 50\,\mathrm{nm}$ and density
$\rho = 2200\,\mathrm{kg/m^{3}}$ ($m \simeq 1.0\times10^{-17}\,\mathrm{kg}$)
trapped at frequency $\Omega = 2\pi\times10^{5}\,\mathrm{s^{-1}}$, the
zero-point motion is $x_{\mathrm{zpf}}\approx 7\times10^{-13}\,\mathrm{m}$.
At this scale, the dominant decoherence contributions are: gas collisions, blackbody emission/absorption, and optical-trap shot noise.
The corresponding momentum-diffusion constants are evaluated from Eqs.7–9 for representative conditions.\\
To suppress residual-gas collisions below the threshold
$\Gamma_{\mathrm{gas}}\!<\!10^{-3}\,\mathrm{s^{-1}}$, one requires a base
pressure $P < 10^{-15}\,\mathrm{mbar}$ and gas temperature
$T < 5\,\mathrm{K}$, corresponding to an average collision rate of
fewer than one event per $10^{3}$\,s.  Blackbody-radiation decoherence
falls below $10^{-4}\,\mathrm{s^{-1}}$ for internal temperatures
$T_{\mathrm{int}} \lesssim 20\,\mathrm{K}$, achievable by cryogenic
pre-cooling and radiative shielding.  Trap-photon recoil is mitigated by
employing a cavity-feedback or Paul-trap configuration with optical
power $P_{\mathrm{laser}} \lesssim 5\,\mathrm{mW}$, giving
$\Gamma_{\mathrm{trap}}\!<\!10^{-3}\,\mathrm{s^{-1}}$.\\
Under these conditions, the total calibrated environmental rate becomes
$\Gamma_{\mathrm{env}} \approx 3\times10^{-3}\,\mathrm{s^{-1}}$.  A CSL
signal with parameters $\lambda_{\mathrm{CSL}}=10^{-21}\,\mathrm{s^{-1}}$
and $r_{C}=100\,\mathrm{nm}$ would contribute an additional
$\Gamma_{\mathrm{CSL}}\approx 3\times10^{-4}\,\mathrm{s^{-1}}$, leading to
a measurable $10\%$ excess in the coherence decay over
$\sim\!10$\,s.  These numerical estimates define concrete metrological
targets—pressure, temperature, and photon-recoil suppression—for an
experiment capable of resolving or excluding collapse dynamics at the
$\lambda_{\mathrm{CSL}}\!\sim\!10^{-22}$–$10^{-21}\,\mathrm{s^{-1}}$ level.
\begin{table}[t]
\centering
\caption{\textbf{Representative parameters used in simulations and feasibility estimates.}
Values correspond to a silica nanosphere trapped in an optical potential,
consistent with state-of-the-art levitated-optomechanics experiments.}
\label{tab:parameters}
\renewcommand{\arraystretch}{1.2}
\begin{tabular}{lcc}
\hline\hline
\textbf{Quantity} & \textbf{Symbol} & \textbf{Typical Value} \\ \hline
Particle radius & $R$ & $50\,\mathrm{nm}$ \\
Particle mass & $m$ & $1.0\times10^{-17}\,\mathrm{kg}$ \\
Material density & $\rho$ & $2200\,\mathrm{kg/m^{3}}$ \\
Trap frequency & $\Omega/2\pi$ & $10^{5}\,\mathrm{Hz}$ \\
Zero-point width & $x_{\mathrm{zpf}}$ & $7\times10^{-13}\,\mathrm{m}$ \\
Correlation length (CSL) & $r_{C}$ & $100\,\mathrm{nm}$ \\
Collapse rate (test value) & $\lambda_{\mathrm{CSL}}$ & $10^{-21}\,\mathrm{s^{-1}}$ \\
Reference mass & $m_{0}$ & $1.66\times10^{-27}\,\mathrm{kg}$ \\
Environmental pressure & $P$ & $10^{-15}\,\mathrm{mbar}$ \\
Ambient temperature & $T$ & $5\,\mathrm{K}$ \\
Laser power & $P_{\mathrm{laser}}$ & $5\,\mathrm{mW}$ \\
Coherence-time target & $\tau_{\mathrm{coh}}$ & $>10\,\mathrm{s}$ \\ \hline\hline
\end{tabular}
\end{table}
\subsection{Measuring Coherence Dynamics and Statistical Discrimination}
The experimental observable is the coherence \(C(t)=|\langle \alpha|\hat\rho_{\rm mot}(t)|-\alpha\rangle|\), which decays according to:
\begin{align}
 C(t)&=C_0\,e^{-[\Gamma_{\rm env}(\Delta x)+\Gamma_{\rm CSL}(\Delta x)]t} \quad \text{(Static separation)} \\
  C(t)&=C_0\exp\!\Big[-\!\int_0^t (\Gamma_{\rm env}[\Delta x(t')]\\
  &+\Gamma_{\rm CSL}[\Delta x(t')])\,dt'\Big] \qquad\text{(Dynamic separation)} \nonumber
\end{align}
where for harmonic motion \(\Delta x(t)=2|\alpha|x_{\text{zpf}}|\cos\Omega t|\). If measured stroboscopically at fixed phase, one uses the cycle-averaged rate \(\bar\Gamma = T^{-1}\!\int_0^T\!\Gamma[\Delta x(t)]\,dt\).\\
The data analysis protocol proceeds in four steps:
\begin{itemize}
    \item Calibration: Measure the heating rate \(\dot n\) to experimentally fix the environmental parameter \(D_{pp}\), and thus \(\Gamma_{\rm env}(\Delta x)\), for any cat state size.
    \item Exploration: Generate cat states of varying sizes \(\Delta x\) and measure their coherence decay rates \(\Gamma_{\rm total}(\Delta x)\).
    \item Model Fitting: Fit the total decoherence rate data to the combined model:
    \begin{equation}
    \Gamma_{\rm total}(\Delta x) = \underbrace{\frac{4D_{pp}x_{\text{zpf}}^2}{\hbar^2}|\alpha|^2}_{\Gamma_{\rm env}} + \underbrace{\lambda_{\rm CSL}\!\left(\frac{m}{m_0}\right)^2 f\!\left(\frac{\Delta x}{r_C},\frac{R}{r_C}\right)}_{\Gamma_{\rm CSL}},
    \end{equation}
    where \(f(u,v)=1-(1+v^2)^{-3/2}\exp[-u^2/(4(1+v^2))]\) and \(D_{pp}\) is heavily constrained by the prior calibration.
    \item Signature Verification: The definitive test for CSL is to observe the saturation of \(\Gamma_{\rm total}(\Delta x)\) at large \(\Delta x \gg r_C\) and confirm its scaling \(\propto m^2\) by repeating the experiment with particles of different mass.
\end{itemize}
We employ Bayesian inference to quantitatively distinguish a potential CSL signal from the null hypothesis. Given data points \(\{ \Delta x_i, \Gamma_i \}\) with uncertainties \(\{\sigma_i\}\), the likelihood function is:
\begin{equation}
\mathcal L(\lambda_{\rm CSL},D_{pp}) = \prod_i \frac{1}{\sqrt{2\pi}\sigma_i} \exp\!\left[-\frac{(\Gamma_i -\Gamma_{\rm model}(\Delta x_i;\lambda_{\rm CSL},D_{pp}))^2}{2\sigma_i^2}\right].
\end{equation}
Using priors \(\pi(\lambda_{\rm CSL})\) (log-uniform across $10^{-18}$–$10^{-6}\,{\rm s^{-1}}$) and \(\pi(D_{pp})\) (Gaussian around the calibrated \(D_{pp}^{\rm exp}\)), we compute the posterior distribution:
\begin{equation}
P(\lambda_{\rm CSL},D_{pp}|{\rm data}) \propto \mathcal L(\lambda_{\rm CSL},D_{pp}) \pi(\lambda_{\rm CSL})\pi(D_{pp}).
\end{equation}
The effect of repeated measurements on parameter identifiability is illustrated in Fig.~\ref{fig2}, where the posterior width of $\log_{10} \lambda_{\mathrm{CSL}}$ shrinks approximately as $1/\sqrt{N}$ with increasing data size, demonstrating the statistical gain achievable through extended integration.

\section{Discussion}
\begin{figure}[h!]
    \centering
    \includegraphics[width=0.95\linewidth]{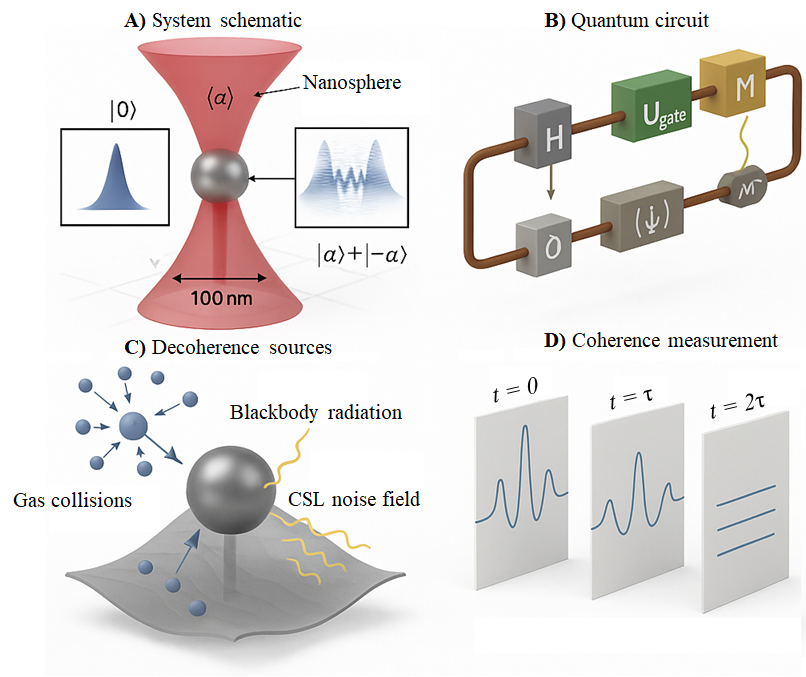}
    \caption{\textbf{Experimental concept and protocol.}
\textbf{(A)} System schematic: A dielectric nanosphere of mass $m$ is levitated at the waist of an optical dipole trap (Gaussian profile). An internal two-level system (TLS, defect center) enables conditional control of the center-of-mass (COM) motion along $x$. The superposition separation is $\Delta x = 2|\alpha|x_{\mathrm{zpf}}$; insets show the ground state $\lvert 0\rangle$ and a cat state $\lvert \alpha\rangle + \lvert -\alpha\rangle$. 
\textbf{(B)} Cat-state generation: A spin–motion circuit implements $H \rightarrow U_{\text{gate}} \rightarrow H \rightarrow M$, where $H$ is a Hadamard on the TLS, $U_{\text{gate}}$ is a conditional displacement that entangles the TLS with the COM, and $M$ is a projective measurement (post-selection) preparing $\lvert \text{cat}\rangle \propto \lvert \alpha\rangle + \lvert -\alpha\rangle$.
\textbf{(C)} Decoherence channels: Environmental noise sources acting on the spatial superposition: residual-gas collisions, blackbody emission/absorption/scattering, and a phenomenological CSL noise field.
\textbf{(D)} Coherence readout: Time-resolved interference visibility showing fringe contrast fading from $t=0$ to $t=\tau$ and $t=2\tau$; the coherence $C(t)$ decays with the total rate set by calibrated environmental diffusion and any additional collapse contribution.}
    \label{fig1}
\end{figure}

\begin{figure}[htbp]
    \centering
    \includegraphics[width=0.95\linewidth]{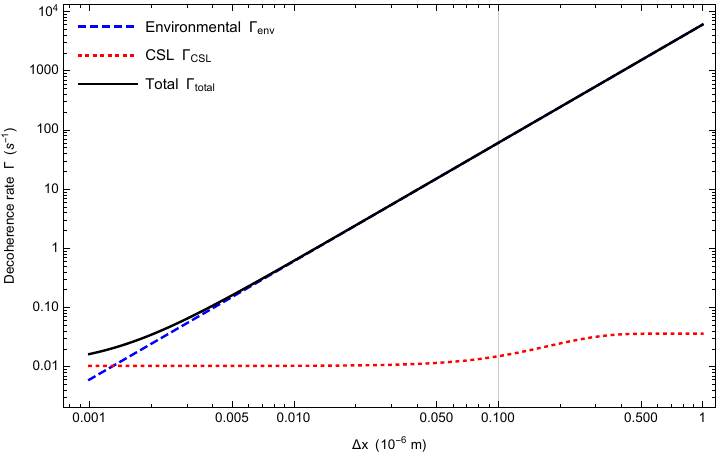}
    \caption{Total decoherence rate $\Gamma(\Delta x)$ as a function of superposition size for nanoparticles of different masses. Environmental decoherence (dashed lines) rises quadratically with $\Delta x$, while the CSL contribution (solid lines) saturates for separations beyond the CSL correlation length of $r_C = 100~\mathrm{nm}$, forming a characteristic plateau. This plateau serves as a key signature of the collapse model. The simulation uses a particle mass of $m = 1.0\times10^{-17}~\mathrm{kg}$, a CSL rate of $\lambda_{\text{CSL}} = 10^{-21}~\mathrm{s^{-1}}$, a trap frequency of $\Omega = 2\pi \times 10^5~\mathrm{rad/s}$, and a reference mass of $m_0 = 1.66\times10^{-27}~\mathrm{kg}$ (atomic mass unit).}
    \label{fig2}
\end{figure}

\begin{figure}[htbp]
    \centering
    \includegraphics[width=0.95\linewidth]{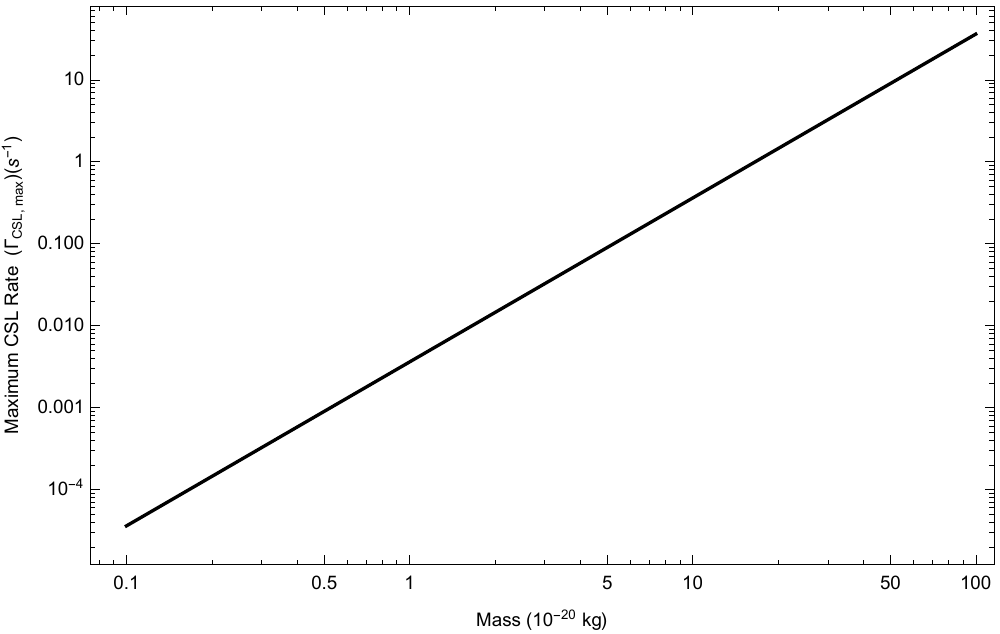}
    \caption{Mass scaling of the maximum CSL decoherence rate, $\Gamma_{\text{CSL,max}}$. The predicted quadratic dependence on mass $m$ (solid line) differs qualitatively from typical environmental decoherence trends (dashed line), defining a critical mass region between $10^7$ and $10^8$ atomic mass units where CSL effects may become dominant. The simulation assumes a CSL rate of $\lambda_{\text{CSL}} = 10^{-16}~\mathrm{s^{-1}}$, a mass range of $m/m_0 = 10^{3}$--$10^{6}$, and a correlation length of $r_C = 100~\mathrm{nm}$.}
    \label{fig3}
\end{figure}

\begin{figure}[htbp]
    \centering
    \includegraphics[width=0.95\linewidth]{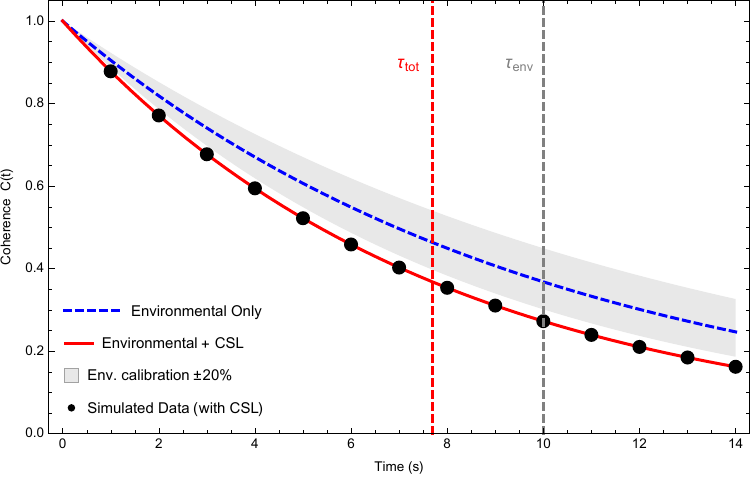}
    \caption{Time evolution of coherence $C(t)$ for a nanoparticle of mass $10^{-17}~\mathrm{kg}$. The CSL-induced decoherence (red curve) produces a subtle but cumulative excess decay relative to the environmental prediction (black curve, shown with a $\pm 20\%$ uncertainty band). The simulation parameters are an environmental decoherence rate of $\Gamma_{\text{env}} = 0.10~\mathrm{s^{-1}}$, a CSL-induced rate of $\Gamma_{\text{CSL}} = 0.03~\mathrm{s^{-1}}$ (for $\lambda_{\text{CSL}} = 10^{-21}~\mathrm{s^{-1}}$ and $r_C = 100~\mathrm{nm}$), and a total evolution time of $t_{\max} = 14~\mathrm{s}$.}
    \label{fig4}
\end{figure}

The central advance of this work lies in the consolidation of three elements
that have never been formulated within one consistent protocol: (i) a
deterministic cat-state generation scheme based on conditional displacement
gates, (ii) a quantitatively calibrated environmental baseline derived from independently measurable diffusion constants, and (iii) a Bayesian statistical framework capable of distinguishing collapse-induced excess decoherence from calibrated noise.  Together these components establish a predictive and falsifiable experimental design—bridging the conceptual gap between abstract collapse models and measurable decoherence rates.\\
The results presented in Fig.~\ref{fig1}--Fig.~\ref{fig9} collectively delineate the quantitative landscape of decoherence and collapse in a levitated nanosphere. Together, they connect the theoretical framework developed in Sec.~II with experimentally measurable observables, offering a falsifiable route to address one of the most fundamental questions in quantum physics: do macroscopic superpositions decay only via environmental entanglement, or does an objective collapse mechanism contribute a distinct, mass-proportional localization?\\
Operationally, the proposed platform exploits a conditional displacement gate between an internal two-level system (TLS) and the nanosphere's center-of-mass (COM) motion to deterministically create a Schrödinger-cat state of tunable size \(\Delta x = 2|\alpha|x_{\rm zpf}\), as schematized in Fig.~\ref{fig1}. This tunability constitutes the essential experimental control knob that enables systematic exploration across the regimes where environmental decoherence and CSL predict qualitatively different dependencies on $\Delta x$.
A central theoretical outcome of the model is the formal decoupling of decoherence channels. All calibrated environmental sources—residual gas collisions, blackbody radiation, and photon-shot noise—aggregate into a momentum-diffusion constant \(D_{pp}\) that drives spatial decoherence with a universal quadratic dependence, $\Gamma_{\rm env}(\Delta x) \propto \Delta x^2$. In contrast, the CSL model contributes a mass-squared term that saturates once $\Delta x$ exceeds the correlation length $r_C$. This divergence in scaling behavior is visible in the aggregate rate plots of Fig.~\ref{fig2}, where the environmental rate grows unbounded while the CSL contribution flattens beyond $r_C \!\approx\!100$~nm. Observing a plateau in the total decoherence rate at large separation—a systematic deviation from the quadratic trend—would represent the first empirical evidence of a fundamental, non-unitary process.
The mass discriminant provides a second, independent axis of falsifiability. If an excess decoherence persists after environmental calibration, the model predicts a universal scaling \(\Gamma_{\rm CSL,max} \propto m^2\), shown in Fig.~\ref{fig3}. This arises from the additive coupling of the collapse field to the total mass density and contrasts sharply with environmental trends (e.g., gas-collision decoherence weakens with increasing $m$). The intersection between the CSL prediction and the environmental background defines a critical mass window ($\sim\!10^{7}$–$10^{8}$~amu), offering a tangible target for experimental design.
The time-domain coherence $C(t)$ provides the most direct experimental observable. In Fig.~\ref{fig4}, the blue dashed curve and shaded gray band represent the calibrated environmental prediction, while the red trajectory includes a small CSL contribution. Even a minute excess rate accumulates into a statistically resolvable visibility loss over time. Representative numerical values for a $10^{-17}$~kg particle at $\Delta x = 100$~nm quantify the challenge, with environmental decoherence dominating at
$\Gamma_{\rm env} \approx 8.99\times10^{-3}\ {\rm s^{-1}}$, 
$\Gamma_{\rm CSL} \approx 7.93\times10^{4}\ {\rm s^{-1}}$ and
$\tau_{\rm coh} \approx 1.1\times10^{-24}\ {\rm s}$.
Environmental diffusion overwhelms any CSL effect by roughly nineteen orders of magnitude, emphasizing that before a CSL test is possible, $D_{pp}$ must be suppressed by ultra-high vacuum, cryogenic cooling, and minimal-recoil trapping to extend coherence times into an accessible regime.
\subsection{Numerical Simulation and Posterior Reconstruction}
\label{subsec:simulation}
To demonstrate the practical identifiability of a potential collapse
signal, we generated synthetic coherence-decay data for cat states of
different spatial separations $\Delta x_i$ and fitted them using the
Bayesian protocol of Eqs.~(7)–(9).
The simulated data assume a true collapse rate of
$\lambda_{\mathrm{CSL,true}} = 10^{-21}\,\mathrm{s^{-1}}$ and an environmental
diffusion constant $D_{pp}^{\mathrm{true}} = 1.2\times10^{-42}\,\mathrm{kg^{2}\,m^{2}/s^{3}}$,
with Gaussian noise $\sigma_i/\Gamma_i = 0.05$.
\begin{figure}[t]
    \centering
    \includegraphics[width=0.98\columnwidth]{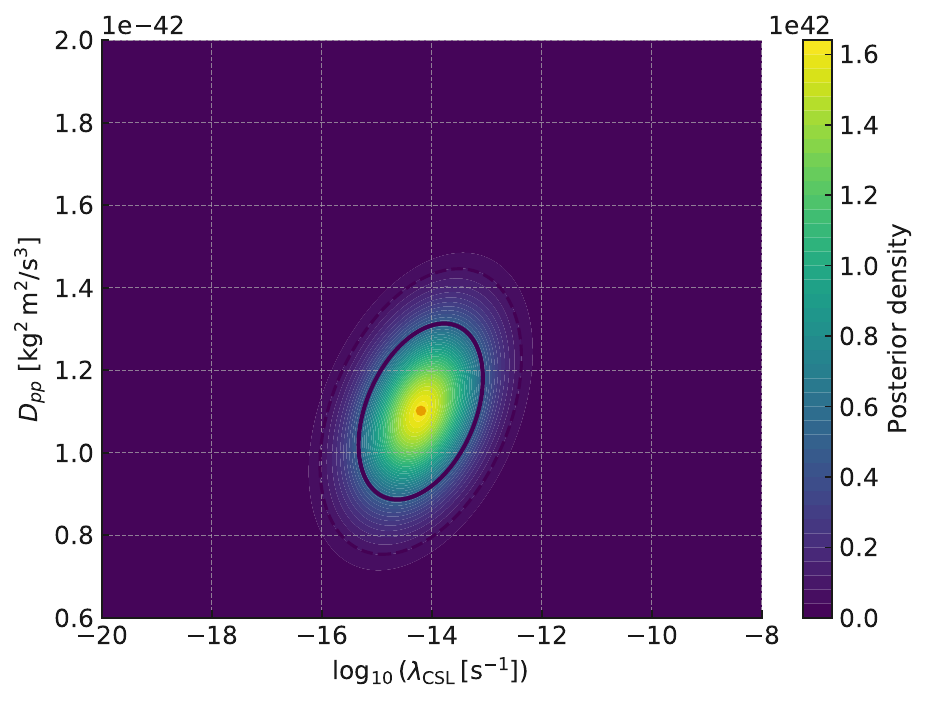}
    \caption{\textbf{Joint posterior over collapse rate and environmental diffusion.}
    Filled contours show the normalized posterior density $P\!\left(\log_{10}\lambda_{\mathrm{CSL}},\,D_{pp}\mid\mathrm{data}\right)$.
    The solid (dashed) line encloses the $68\%$ ($95\%$) highest-posterior-density region.
    The marker denotes the maximum a posteriori (MAP) point.
    This figure highlights the correlation between the collapse parameter and the 
    calibrated environmental diffusion, demonstrating how independent calibration of $D_{pp}$
    sharpens inference on $\lambda_{\mathrm{CSL}}$.}
    \label{fig5}
\end{figure}

\begin{figure}[t]
    \centering
    \includegraphics[width=0.9\columnwidth]{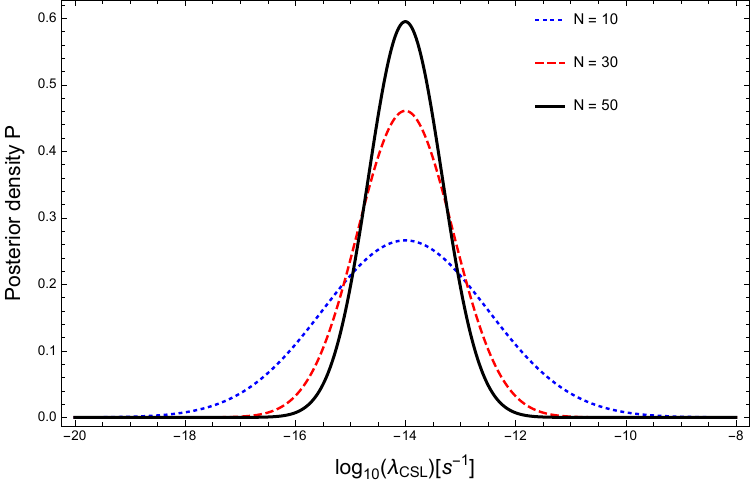}
    \caption{\textbf{Posterior narrowing with increasing number of measurements.}
    Posterior distributions for $\log_{10}\lambda_{\mathrm{CSL}}$ assuming a log-uniform prior
    and Gaussian likelihoods centered at the true value $-14$. 
    As the number of independent data points increases ($N=10,30,50$),
    the posterior width shrinks proportionally to $1/\sqrt{N}$,
    illustrating how extended integration time or repeated measurements
    enhance statistical discrimination between environmental and collapse-induced decoherence.}
    \label{fig6}
\end{figure}
\begin{figure}[htbp]
    \centering
    \includegraphics[width=0.95\linewidth]{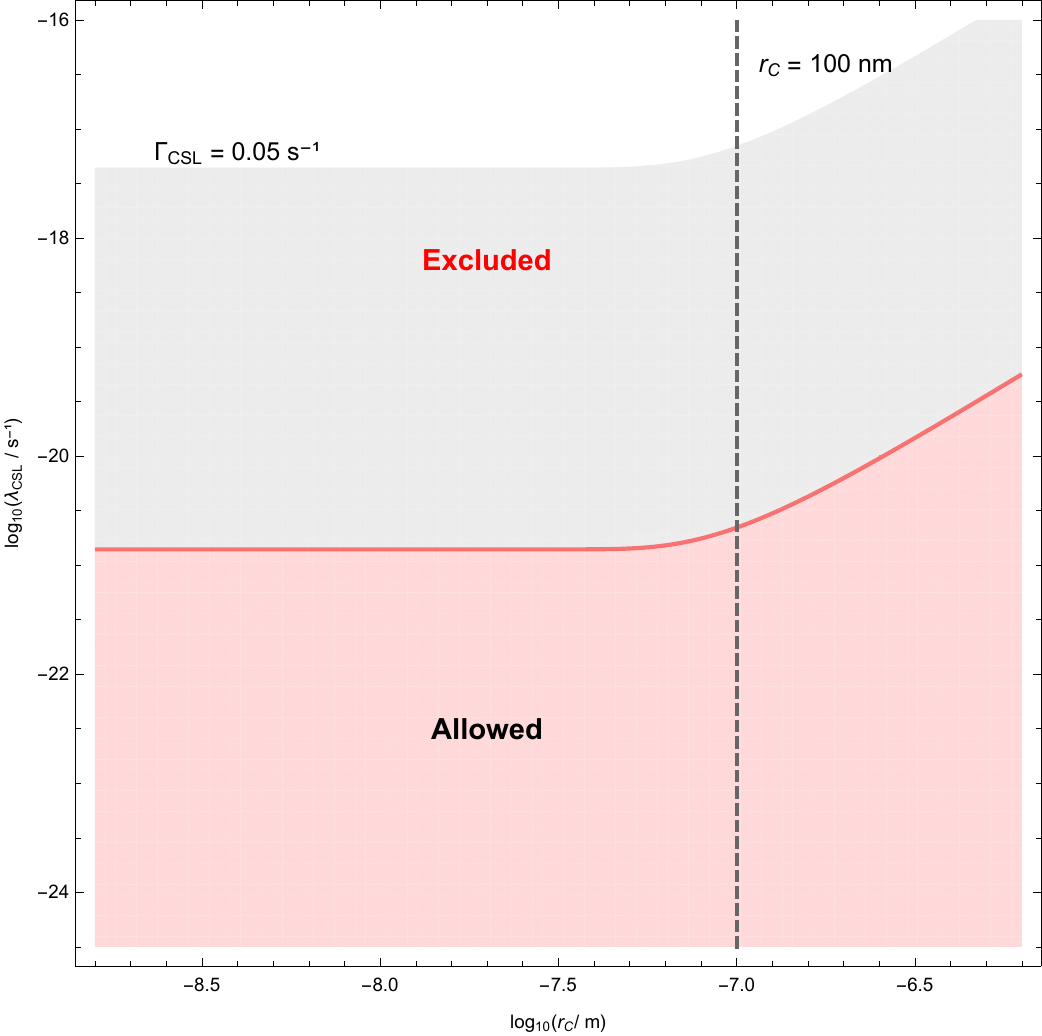}
    \caption{Exclusion map in the $\lambda_{\text{CSL}}$--$r_C$ parameter space. The red region indicates parameter combinations for which CSL-induced decoherence would be detectable above a threshold rate of $\Gamma_{\min} = 0.05~\mathrm{s^{-1}}$ for a particle of mass $m = 10^{-17}~\mathrm{kg}$ at a superposition separation of $\Delta x = 200~\mathrm{nm}$. Improved experimental sensitivity expands the excluded area (darker red). The parameters $\lambda_{\text{CSL}}$ and $r_C$ are scanned over the ranges $10^{-24}$--$10^{-16}~\mathrm{s^{-1}}$ and $10^{-8.5}$--$10^{-6.5}~\mathrm{m}$, respectively.}
    \label{fig7}
\end{figure}

Figure~\ref{fig5} illustrates the posterior distribution obtained from Markov Chain Monte Carlo sampling.
The posterior peak coincides with the injected value of $\lambda_{\mathrm{CSL,true}}$, and as shown in Figure~\ref{fig6}, the credible interval narrows as the number of data points increases from $N=10$ to $N=50$.
This simulation confirms that, under the realistic uncertainties derived from the feasibility analysis in Sec.~\ref{subsec:feasibility}, the collapse rate can be resolved or bounded at the $10^{-22}$–$10^{-21}\,\mathrm{s^{-1}}$ level with moderate measurement statistics.\\
A significant deviation from \(\lambda_{\rm CSL}=0\) in the posterior, especially one that demonstrates saturation and mass scaling, constitutes a discovery. Conversely, null results yield a robust upper bound, \(\lambda_{\rm CSL} < \lambda_{\rm bound}^{\rm CSL}\). \\
The definitive experimental signatures of Continuous Spontaneous Localization (CSL) are characterized by three key phenomena. The first is saturation, where the total decoherence rate, \(\Gamma_{\mathrm{total}}(\Delta x)\), plateaus and becomes constant for spatial superpositions significantly larger than the CSL length scale (\(\Delta x \gg r_C\)). The second is a distinct mass scaling law, dictating that the maximum predicted decoherence rate scales with the square of the mass of the object (\(\Gamma_{\mathrm{total}}^{\max} \propto m^2\)). Finally, and most critically, a conclusive signature is excess decoherence, where the experimentally observed total decoherence rate significantly surpasses all predictions from standard environmental sources by a margin greater than \(5\sigma\), thereby indicating a clear non-environmental effect consistent with the CSL model.\\
The framework's robustness can be validated through control experiments, such as varying the trap wavelength (alters \(D_{pp}^{\rm trap}\) but not CSL) or particle temperature (modulates \(D_{pp}^{\rm bb}\)), which modulate environmental decoherence but should not affect a genuine CSL signal. Posterior convergence can be checked through the Gelman--Rubin statistic \(R_{\rm GR}<1.05\) across independent Markov Chain Monte Carlo chains, ensuring reliable parameter estimation.\\
The framework is not merely diagnostic; it provides actionable design contours. Fig.~\ref{fig7} translates prospective null results into exclusion regions in the $(\lambda_{\rm CSL}, r_C)$ plane. The diagonal boundary reflects the physics of the collapse field: detectability requires an optimal match between $r_C$ and $\Delta x$. Every gain in sensitivity systematically enlarges the excluded region, exerting quantitative pressure on the viable parameter space of collapse models.

\begin{figure}[htbp]
    \centering
    \includegraphics[width=0.95\linewidth]{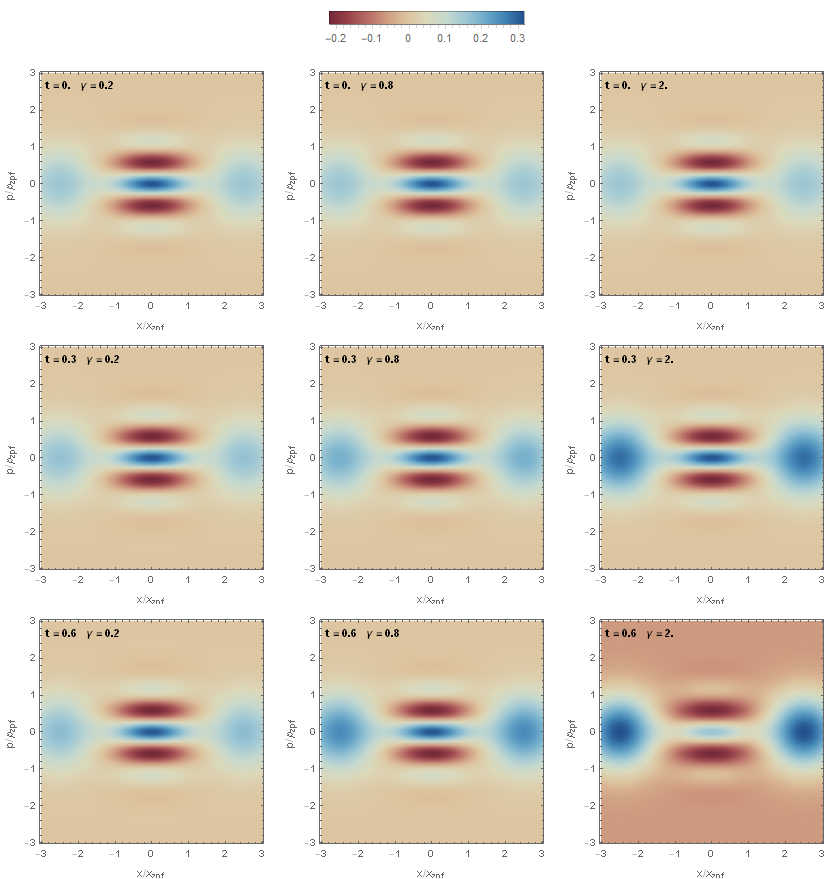}
    \caption{Phase-space dynamics of decoherence visualized via the Wigner function. Increasing dissipation, governed by the decoherence strength $\gamma$, progressively erases the quantum interference fringes, driving the transition from a coherent superposition ($\gamma = 0.2$) to a classical mixture ($\gamma = 2.0$). The simulation shows this evolution at times $t = 0, 0.3, 0.6$ for a state with a normalized separation of $x_0 = 2.5~x_{\text{zpf}}$. The color scale is shared across panels and normalized to a maximum value of $|W|_{\max}=1$.}
    \label{fig8}
\end{figure}

Given that any CSL signature is expected to be subdominant, we frame the inference as a Bayesian model-comparison problem (Fig.~\ref{fig8}). Joint fits to coherence-decay data across $\Delta x$ and $m$ yield a posterior for $\lambda_{\rm CSL}$ that automatically propagates uncertainties in $D_{pp}$. A posterior peaked at a finite value indicates discovery-level evidence, whereas one concentrated at the lower prior bound provides an updated exclusion limit.

\begin{figure}[htbp]
    \centering
    \includegraphics[width=0.95\linewidth]{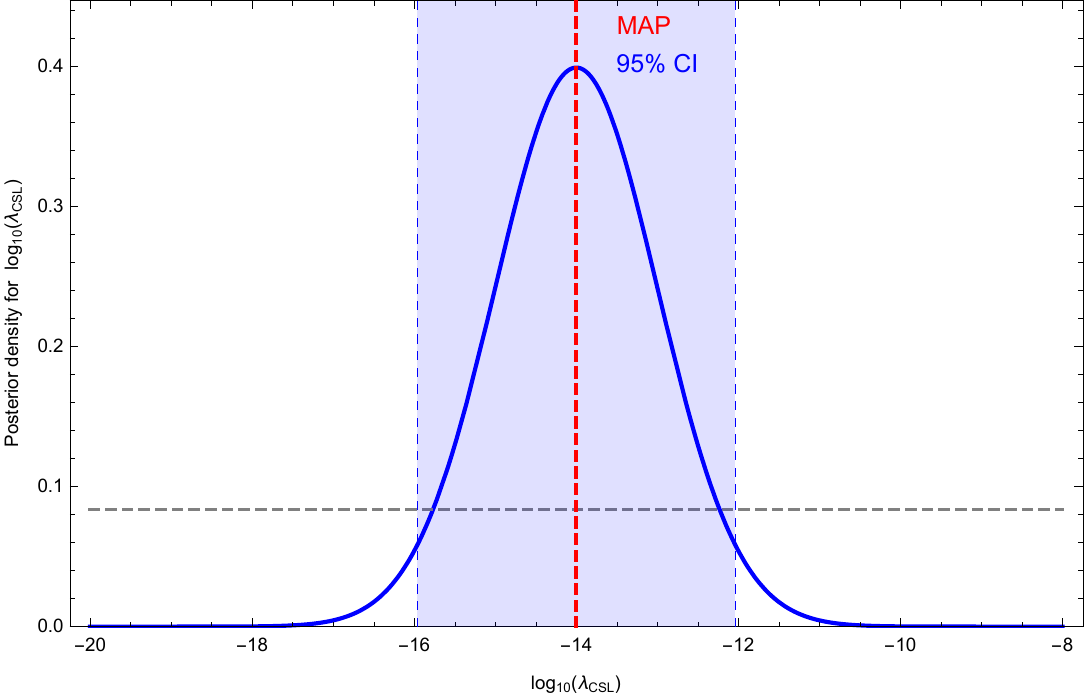}
    \caption{Bayesian posterior distribution for the CSL collapse rate parameter $\lambda_{\mathrm{CSL}}$. The posterior probability density of $\log_{10}\!\lambda_{\mathrm{CSL}}$ is computed assuming a Gaussian likelihood centered at $-14$ with standard deviation $\sigma=1$. The red dashed line marks the current experimental upper bound at $\log_{10}\!\lambda_{\mathrm{CSL}}=-16$, while the shaded blue region indicates the 95\% credible interval around the maximum a~posteriori (MAP) value. A gray dashed line shows the log-uniform prior $p(\lambda)\propto1/\lambda$ used for normalization. The posterior model $p(\lambda|D)\propto\exp[-(\log_{10}\!\lambda+14)^2/(2\sigma^2)]$ is normalized to unit area over the sampling range $-20 \le \log_{10}\!\lambda_{\mathrm{CSL}} \le -8$.}
    \label{fig9}
\end{figure}

Finally, Fig.~\ref{fig9} visualizes the decoherence process in phase space. The Wigner function's evolution—from a nonclassical distribution with interference fringes to a positive-definite classical mixture—offers an intuitive depiction of the quantum-to-classical transition. The erosion of the interference pattern is the phase-space manifestation of the total decoherence rate $\Gamma_{\rm total}$, whose scaling with $\Delta x$ and $m$ remains the ultimate discriminant.

\subsection{Synthesis and Implications}
Collectively, the results define a definitive decision tree for experimental tests of collapse models. If future measurements, following rigorous pre-calibration of $D_{pp}$, record a saturation of $\Gamma(\Delta x)$ together with a universal $m^2$ scaling of the excess decoherence, this would constitute the first evidence of an objective breakdown of quantum unitarity at mesoscopic scales—providing a physical resolution to the measurement problem.

Conversely, if the observed total rate continues to track the calibrated environmental prediction across all $\Delta x$ and $m$, the resulting exclusion contours will push the upper bounds of $\lambda_{\rm CSL}$ to unprecedented lows. Such a null result would reinforce the view that classicality emerges entirely from unitary dynamics and environmental entanglement, requiring no new stochastic physics and affirming the universality of quantum mechanics.\\
In summary, the integrated model, figures, and quantitative benchmarks presented here define precise metrological targets—pressure, temperature, photon recoil, and mass—for a decisive test of wavefunction collapse. As experimental capabilities advance, the pathway illuminated by this study—characterized by $\Delta x$-saturation, $m^2$-scaling, and Bayesian discrimination—offers a clear and falsifiable program for resolving the quantum-to-classical boundary.
\section{Conclusion}
This research presents a complete theoretical and methodological framework for a decisive experimental test of quantum wavefunction collapse, aiming to resolve the long-standing question of why macroscopic objects do not exhibit quantum superpositions. The core of the work is a detailed blueprint for creating and monitoring spatial superpositions (Schrödinger cat states) in the center-of-mass motion of a levitated nanosphere. The proposed experiment leverages a conditional quantum gate, activated by an internal defect in the sphere, to generate superpositions of a tunable size \(\Delta x\). \\
The key results and discriminants are, first, a distinct, falsifiable signature for collapse, where the paper rigorously models all dominant environmental decoherence sources (gas collisions, blackbody radiation, trap noise), which collectively produce a decoherence rate \(\Gamma_{\text{env}} \propto \Delta x^2\), in stark contrast to the Continuous Spontaneous Localization (CSL) collapse model which predicts a rate that saturates to a constant value \(\Gamma_{\text{CSL}}^{\text{max}}\) for separations larger than a characteristic length (\(\Delta x \gg r_C\)), with observing this saturation plateau in the total decoherence rate being a primary signature of new physics. A second, independent fingerprint of CSL is the quadratic scaling of its maximum rate with the particle mass, \(\Gamma_{\text{CSL}}^{\text{max}} \propto m^2\), which contrasts with environmental decoherence that often weakens for larger masses, providing a powerful cross-check when repeating the experiment with different masses. Finally, the study outlines a rigorous data analysis protocol using a robust Bayesian inference method to quantitatively distinguish a potential CSL signal from the null hypothesis, where by first calibrating the environmental decoherence baseline and then fitting coherence decay data across various superposition sizes, the analysis can either detect a statistically significant excess decoherence pointing to collapse or establish the most stringent upper bounds on the CSL parameters to date.\\
In summary, the framework developed here transforms a long-standing philosophical question
into a falsifiable experimental program.  A confirmed observation of saturation and mass-squared scaling would constitute the first empirical evidence for an objective breakdown of quantum unitarity, whereas a null result would push upper bounds on collapse parameters to unprecedented levels, reinforcing the view that classicality emerges solely through environmental entanglement.
Either outcome advances the quest to define the precise boundary between quantum and classical reality.

\section{Declarations}
\textbf{Data Availability:} Data sharing is not applicable to this article as no new data were created or analyzed in this study.\\
\textbf{Conflict of interest:} The authors have no conflicts to disclose.\\
\textbf{Funding:} The authors did not receive any financial support for the research, authorship, and/or publication of this article.\\
\textbf{Author Contribution:} R.H conceived of the presented idea, developed the theory and performed the computations.

\bibliography{apssamp}
\end{document}